\documentstyle[aps,preprint,tighten,graphicx]{revtex}

\begin{document}

% A useful Journal macro
\def\Journal#1#2#3#4{{#1} {\bf #2}, #3 (#4)}

% Some useful journal names
\def\NP{{\em Nucl. Phys.} }
\def\NC{{\em Nuovo Cim.} }
\def\NCL{{\em Nuovo Cim. Lett.} }
\def\PL{{\em Phys. Lett.} }
\def\PR{{\em Phys. Rev.} }
\def\PRL{{\em Phys. Rev. Lett.} }
\def\PREP{{\em Phys.  Rep.} }
\def\AP{{\em Ann. of Phys.} }
\def\ZP{{\em Z. Phys.} }
\def\SJNP{{\em Sov. J. Nucl. Phys.} }
\def\JETPL{{\em Jrn. of Exp. and Theo. Phys. Lett.} }
\def\JPSJ{{\em Jrn. Phys. Soc. Jpn.} }
\def\EPJ{{\em The Europ. Phys. Jrn.} }
\def\PN{{\em PiN-Newsletter} }

\newcommand{\mathbold}[1]{\mbox{\protect\boldmath $\displaystyle #1$}}

\def\rme{{\rm e}}
\def\rmi{{\rm i}}
\def\rmx{{\rm x}}

\def\intm1p1{{\int\limits_{-1}^{1}}}
\def\sumlUNE{{\sum\limits_{l=0}^{\infty}}}

\newcommand{\einh}{\frac{1}{2}}
\newcommand{\dreih}{\frac{3}{2}}
\newcommand{\eind}{\frac{1}{3}}
\newcommand{\zweid}{\frac{2}{3}}
\newcommand{\einse}{\frac{1}{6}}

\newcommand{\be}{\begin{equation}}
\newcommand{\ee}{\end{equation}}
\newcommand{\bea}{\begin{eqnarray}}
\newcommand{\eea}{\end{eqnarray}}

\newcommand{\nres}[4]{#1_{#2#3}(#4)}

\newcommand{\lmnr}[2]{{\cal L}_{\varphi N #1}^{#2}}

\newcommand{\real}{{\rm Re}}
\newcommand{\imag}{{\rm Im}}

\newcommand{\bsl}[1]{#1 \!\!\! /}

\newcommand{\abs}[1]{\left | #1 \right |}

\newcommand{\dop}[1]{#1^{\prime}}
\newcommand{\dops}[1]{#1^{\prime \, 2}}
\newcommand{\dopp}[1]{#1^{\prime \prime}}
\newcommand{\hatdop}[1]{\hat #1^{\, \prime}}

\newcommand{\Iba}{b a}
\newcommand{\Ipipi}{\pi \pi}
\newcommand{\Ipigam}{\pi \gamma}
\newcommand{\Ibgam}{b \gamma}
\newcommand{\Iphigam}{\varphi \gamma}
\newcommand{\Iphipgam}{\dop \varphi \gamma}
\newcommand{\Iphiphip}{\varphi \dop \varphi}
\newcommand{\Igamgam}{\gamma \gamma}

\newcommand{\solid}{\protect\rule[1mm]{6mm}{.1mm}}
\newcommand{\dash}{\protect\rule[1mm]{2.5mm}{.1mm}\hspace{1mm}
\protect\rule[1mm]{2.5mm}{.1mm}}
\newcommand{\shortdash}{\protect\rule[1mm]{1mm}{.1mm}\hspace{4mm}
\protect\rule[1mm]{1mm}{.1mm}}
\newcommand{\dashdot}{\protect\rule[1mm]{2.5mm}{.1mm}\hspace{1mm}$\cdot\cdot$}
\newcommand{\dashsdot}{\protect\rule[1mm]{2.5mm}{.1mm}\hspace{1mm}$\cdot$}
\newcommand{\dotdot}{\protect$\cdot\!\cdot\!\cdot$}

\newcommand{\FullBox}{\protect\rule[0.5mm]{2.5mm}{2.5mm}}

\newcommand{\knb}{\overline{K}^0}
\newcommand{\mb}[1]{\overline{#1}}

\newcommand{\Pp}{\pi^+}
\newcommand{\Pn}{\pi^0}
\newcommand{\Pm}{\pi^-}
\newcommand{\Spl}{\Sigma^+}
\newcommand{\Sn}{\Sigma^0}
\newcommand{\Sm}{\Sigma^-}

\pagestyle{plain} \pagenumbering{arabic}

\title{Anti-Kaon Induced Reactions on the Nucleon\footnote{Work
      supported by BMBF and GSI Darmstadt}}

\author{M. Th. Keil \footnote
    {e-mail: Mathias.T.Keil@theo.physik.uni-giessen.de}, G. Penner 
         and U. Mosel}

\address{Institut f\"ur Theoretische Physik, Universit\"at Giessen\\
    D--35392 Giessen, Germany \\[3mm]}
\maketitle

%%%%%%%%%%%%%%%%%%%%%%%%%%%%%%%%%%%%%%%%%%%%%%%%%%%%%%%%%%%%%%%%%%%%%%%%%
%
% Abstract
%
%%%%%%%%%%%%%%%%%%%%%%%%%%%%%%%%%%%%%%%%%%%%%%%%%%%%%%%%%%%%%%%%%%%%%%%%%

\begin{abstract}
Using a previously established effective Lagrangian model we 
describe anti-kaon induced reactions on the nucleon. The dominantly 
contributing channels in the cm-energy region from threshold up to 1.72 
GeV are included ($\overline K N$, $\pi \Sigma$, $\pi \Lambda$). We solve
the Bethe-Salpeter equation in an unitary $K$-matrix approximation. 
\end{abstract}

\clearpage

%%%%%%%%%%%%%%%%%%%%%%%%%%%%%%%%%%%%%%%%%%%%%%%%%%%%%%%%%%%%%%%%%%%%%%%%%
%
% Introduction
%
%%%%%%%%%%%%%%%%%%%%%%%%%%%%%%%%%%%%%%%%%%%%%%%%%%%%%%%%%%%%%%%%%%%%%%%%%

\section{Introduction}

One major goal in hadron physics is to find the properties of
resonances. In the sector of $N^*$ and $\Delta$ resonances there
exists a huge amount of data especially from $\pi N$ reactions. The
properties of most of these resonances are known quite well
\cite{pdb98} from various analyses of the experimental data (see for
example \cite{ms92}) and models that are able to describe these data,
as for example \cite{fm}. 
This situation changes when we look at resonances with strangeness
$-1$. Here the properties of some resonances are not settled at all
(e.g. in the $P_{01}$ partial wave there might be one or two
resonances with barely known masses and widths). In order to learn more
about these resonances and also the coupling constants involving
strange baryons, we have extended the model of \cite{fm} to the
strangeness $-1$ sector (i.e. the asymptotic states are $\overline K
N$, $\pi \Sigma$, and $\pi \Lambda$).

Reference \cite{fm} sets the framework of the present
calculation. Using the $K$ matrix approxi\-mation, $\pi$ and $\gamma$
induced reactions on the nucleon with final states $\pi N$, $\gamma
N$, $\pi \pi N$, $\eta N$, $K \Lambda$, and since last year also $K
\Sigma$, were calculated. Also on the basis of \cite{fm} Bennhold et
al. \cite{wak} are investigating nucleon and $\Delta$ resonances. In
their latest publication \cite{Be00}, special emphasis is put on the
photoproduction of kaon hyperon states. Especially with the inclusion
of the final state $K \Sigma$ it was possible to extend the energy
range up to $\sqrt s = 2$ GeV and thus to investigate the resonances
around 1900 MeV. For the photoproduction it is very important to
maintain gauge invariance after introducing form factors. Using a form
factor prescription by Haberzettl \cite{habfeu} yields an excellent
agreement between experimental data and model calculations.

In general, experimental observables are very
well described with the model in \cite{fm}; however, there is a
problem concerning the final states $K \Lambda$ and $K \Sigma$; the
angular differential cross sections for the $\pi$ induced reactions do
not agree with experiment. In the reactions $\pi N \rightarrow K
\Lambda$ or $K \Sigma$ there are baryons and resonances with
strangeness $-1$ propagating in $u$-channel diagrams. Since these are
nonresonant contributions and the couplings of the baryons and widths
of the resonances are, if at all, poorly known, these diagrams were
not included in \cite{fm}. 

The idea is now to determine these parameters in a calculation where
they contribute in resonant diagrams, which is the case for $\overline
K$ induced reactions on the nucleon.

In this paper we present our first results. Section \ref{model} shows
the concept of our calculation with the unitarity conserving $K$
matrix approximation, which reduces the Bethe-Salpeter equation from a
system of coupled integral equations to the problem of matrix
inversion. Section \ref{lagrden} covers the used Lagrange density and
motivates the final states that have to be considered in the
calculation. In section \ref{ff} the used form factors are given and
in the following section \ref{ex} the experimental data that were used
for the fit are discussed. Our results are presented in section
\ref{res}. Section \ref{cno} contains conclusions and outlook.

%%%%%%%%%%%%%%%%%%%%%%%%%%%%%%%%%%%%%%%%%%%%%%%%%%%%%%%%%%%%%%%%%%%%%%%%%
%
% The Model
%
%%%%%%%%%%%%%%%%%%%%%%%%%%%%%%%%%%%%%%%%%%%%%%%%%%%%%%%%%%%%%%%%%%%%%%%%%

\section{The Model}
\label{model}

\begin{figure}
\begin{center}
\includegraphics[width=12cm]{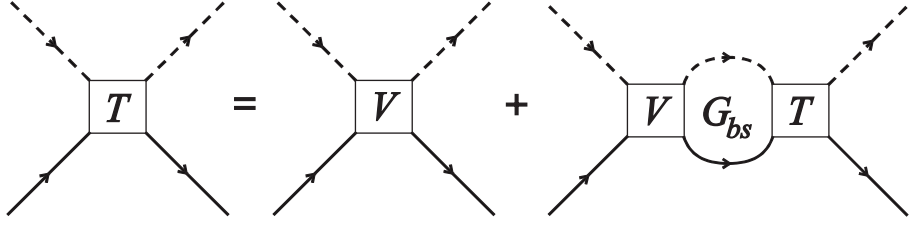}
\end{center}
\caption{Schematic picture of the Bethe Salpeter Equation.}
\label{bse_p}
\end{figure}

All information of the scattering process is contained in the
scattering matrix $S$. This matrix can be decomposed into a trivial
part and the part that includes all information about reactions, the
$T$-matrix
\be
S = I + \rmi \; T .
\ee
The scattering equation that needs to be solved is the Bethe Salpeter 
equation (BSE). Schematically the BSE is given as illustrated in
fig. \ref{bse_p}. $V$ is the amplitude of the elementary interactions
and is calculated using Feynman rules which follow from a
Lagrangian that is given in section \ref{lagrden}. The two-particle
propagator $\mathcal G_{BS}$ is given by the product of the baryon and
the meson propagator. The integration over the loop momentum yields a
system of coupled integral equations for all final states.

We introduce the Feynman amplitude $\mathcal M$
\be
\langle  \mathbold p_{1} \mathbold p_{2} | \rmi \hat T 
| \mathbold   k_{A} \mathbold k_{B} \rangle = 
(2 \pi)^4 
\delta^{(4)}(k_A + k_B - (p_1 + p_2))
\rmi
{\mathcal{M}}( k_A, k_B \rightarrow p_1, p_2).
\ee
Taking out the spinors we define the amplitude $M_{fi}$
\be
{\cal M}_{fi} = \bar u(\dop p, \dop s) M_{fi} u(p,s).
\label{mwithu}
\ee
Then the BSE has the form
\be
M (\dop p, p; \sqrt s) = V (\dop p, p; \sqrt s) +
\int \frac{d^4 k}{(2 \pi)^4} V (\dop p, k; \sqrt s) \;
{\cal G}_{BS} (k; \sqrt s) \;
M (k, p, \sqrt s) .
\label{bs}
\ee

To make this problem solvable we are using a $K$ matrix
approximation. In our notation the $K$ matrix is defined by
\bea
K &=& V (\dop p, p; \sqrt s) + {\mathrm P} \int \frac{d^4 k}{(2 \pi)^4}
V (\dop p, k; \sqrt s) \; {\cal G}_{BS} (k; \sqrt s) \; K (k, p, \sqrt
s) \nonumber \\
&=& V (\dop p, p; \sqrt s) + \int \frac{d^4 k}{(2 \pi)^4} V (\dop
p, k; \sqrt s) \; \Re {\cal G}_{BS} (k; \sqrt s) \; K (k, p, \sqrt s)
,
\label{kteqn1}
\eea
where P stands for the principal value of the integral and $\Re {\cal
  G}_{BS}$ stands for the real part of the propagator. The latter
contains the pole of the propagator as can be seen from the
following. A propagator is given in the form
\be 
{1\over {a + \rmi \epsilon}} = {{a - \rmi \epsilon} \over {a^2 +
 \epsilon^2}} \;\; \stackrel{\epsilon \rightarrow 0}\longrightarrow
\;\; {1 \over a} - \rmi \pi \delta(a) ,
\ee
so the real part contains the pole and the imaginary part is
proportional to a delta function with the on-shell condition as its
argument.

From the unitarity of the $S$-matrix, $S^\dagger S = I$, follows
\be
- \rmi (T - T^\dagger) = T^\dagger T,
\ee
and using the BSE we get 
\be
M - M^* = 
2 \rmi \; M^* \; \Im ({\cal G}_{BS}) \; M .
\label{tunitar}
\ee
Thus, in order to keep unitarity we must preserve the imaginary part
of the propagator. The real part, on the other hand, plays no role for
unitarity, so we simply neglect it
\be
\Re ({\mathcal G}_{BS}) = 0 .
\ee
This is the same as putting the
intermediate particles on their mass shell. In  \cite{fm} and
\cite{pj91} it is shown that this is a reasonable assumption for
energies not too close to a particle production threshold.

With this approximation the propagator $\mathcal G_{BS}$ becomes
\be
{\cal G}_K = - \rmi 2 \pi^2 
\delta (k_B^2 - m_B^2) \delta (k_M^2 - m_M^2)
\theta (k^0_B) \theta (k^0_M) (\bsl k_B + m_B)
\label{kmatprop}
\ee
and from (\ref{kteqn1}) we get 
\be
K=V \; .
\ee

With the delta functions in the propagator the integral in the BSE
(\ref{bs}) is easy to evaluate and the problem reduces to the
inversion of an $n \times n$ matrix, if $n$ is the number of
energetically allowed channels. 

Using
\be 
(\bsl k_B + m_B) = \sum_s u(k_B,s) \bar u(k_B,s)
\ee
and going back to the notation of the amplitudes including the spinors
(\ref{mwithu}) we get
\be
{\mathcal M} = (I - \rmi \; {\mathcal K})^{-1} \; {\mathcal K} ,
\ee
where ${\cal K} = {\cal V} = \bar u V u$.

%%%%%%%%%%%%%%%%%%%%%%%%%%%%%%%%%%%%%%%%%%%%%%%%%%%%%%%%%%%%%%%%%%%%%%%%%
%
% The Lagrange density
%
%%%%%%%%%%%%%%%%%%%%%%%%%%%%%%%%%%%%%%%%%%%%%%%%%%%%%%%%%%%%%%%%%%%%%%%%%

\section{The Lagrange density}
\label{lagrden}

\begin{figure}
\begin{center}
\includegraphics[width=12cm]{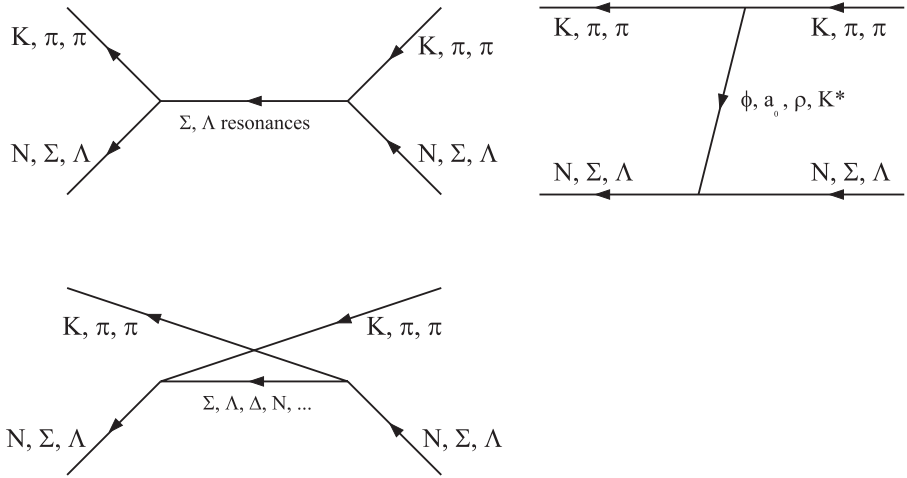}
\end{center}
\caption{Feynman diagrams used in the calculation. In these diagrams
  the time runs from the right to the left, as usual in matrix
  elements.}
\label{diag}
\end{figure}

In order to calculate the entries of the matrix $\cal K$ all
elementary diagrams must be calculated. We restrict ourselves to
tree level diagrams (cf. fig. \ref{diag}) using the physical masses
and charges for the asymptotic particles and $t$-channel mesons. The
parameters of $N$, $\Delta$, and resonances in $u$-channel diagrams are
taken from \cite{fm} and \cite{GP00}.

\begin{figure}
\begin{center}
\includegraphics[width=12cm]{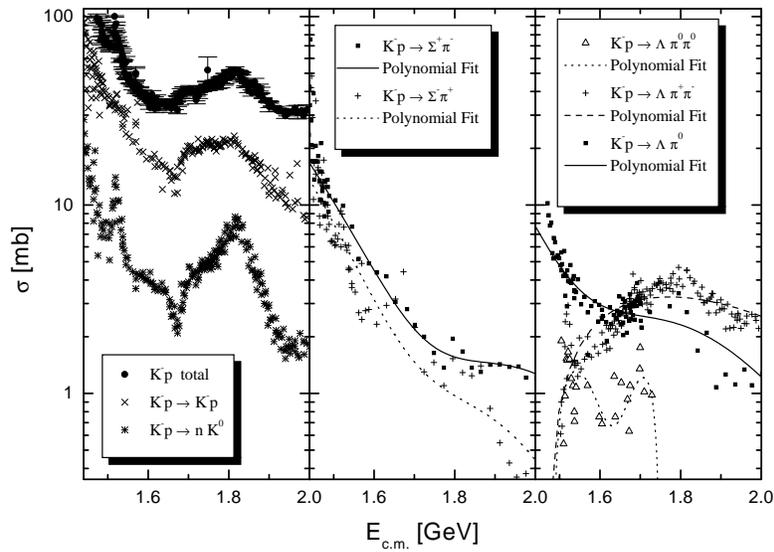}
\end{center}
\caption{Total cross sections}
\label{tox}
\end{figure}

The channels included in the calculation are $\overline K N$, $\pi
\Sigma$, and $\pi \Lambda$. These are the dominant contributions in the
relevant energy interval (fig. \ref{tox}).
Because we have to fix the particle properties in this
phenomenological model with the help of experimental data, we will
include more channels as soon as reliable data are available. 

In order to calculate the $s$-, $t$-, and $u$-channel diagrams for the
relevant channels we are using the following Lagrange density:

\bea 
{\cal L} 
= &&\bar \Psi ( \rmi \bsl \partial - m_\Psi) \Psi +
{1\over2} (\partial_\mu\partial^\mu - m_\varphi^2)\varphi^2 +
{1\over4} v_{\mu\nu}v^{\mu\nu} + {1\over2} v_\mu v^\mu m_v^2
\nonumber \\
&+&\bar \Psi_R ( \rmi \bsl \partial - m_R) \Psi_R +
\bar \Psi_R^\alpha ( \rmi \bsl \partial - m_R) \Psi_{R \alpha}
\nonumber \\ 
&-& \frac{g_{\varphi\Psi\Psi}}{2 m_\Psi} [ \bar \Psi \gamma_5
\gamma_{\mu} (\partial^{\mu} \varphi) \Psi + h.c. ] \nonumber 
\\
&-& g_{v\Psi\Psi} [ \bar \Psi \left ( \gamma_{\mu} v^{\mu} - \kappa_{v
 \Psi \Psi}
\frac{\sigma_{\mu \nu}}{4 m_\Psi} v^{\mu \nu} \right ) \Psi + h.c. ]
\nonumber \\ 
&-& \rmi g_{v\varphi\varphi} \left [ \varphi (\partial_{\mu} \varphi)
 v^{\mu} + h.c. \right ]
\nonumber \\ 
&-& \frac{g_{\varphi\Psi\Psi_R}}{m_R + m_\Psi} [ \bar \Psi_R^-
\gamma_{\mu} (\partial^{\mu} \varphi ) \Psi + h.c. ]
\nonumber \\
&-& \frac{g_{\varphi\Psi\Psi_R}}{m_R - m_\Psi} [ \bar \Psi_R^+
\gamma_{\mu} \gamma_5 (\partial^{\mu} \varphi ) \Psi + h.c. ]
\nonumber \\ 
&-&\frac{g_{\varphi\Psi\Psi_\dreih}}{m_{\pi}} \left [ \bar \Psi_R^{\alpha -}
[ g_{\alpha \mu} - \frac{1}{2} (1 + 2 z_{\varphi \Psi}) \gamma_{\alpha}
\gamma_{\mu} ] \gamma_5 (\partial^{\mu} \varphi) \Psi + h.c. \right ]  
\nonumber \\ 
&-& \frac{g_{\varphi\Psi\Psi_\dreih}}{m_{\pi}} \left [ \bar \Psi_R^{\alpha +}
[ g_{\alpha \mu} - \frac{1}{2} (1 + 2 z_{\varphi \Psi})
\gamma_{\alpha} \gamma_{\mu}] 
(\partial^{\mu} \varphi) \Psi + h.c. \right ]
\label{lagr}
\eea
The kinetic and mass terms are given in the first two lines. $\Psi$
refers to the Baryons (N, $\Sigma$, $\Lambda$), $\varphi$ to
pseudo scalar mesons (K, $\pi$), $v^{\mu}$ stands for the vector mesons
($K^*$, $\phi$, $\rho$, $a_0$), $\Psi_R^\pm$ and $\Psi_R^{\alpha \pm}$
are referring to Spin $\einh$ and $\dreih$ resonances resp. ($\pm$
stands for the parity of the resonance).

Pseudo scalar mesons and the spin $\einh$ baryons are coupled by
pseudo vector coupling terms. Baryons and vector mesons are coupled by
the superposition of vector and tensor coupling terms and $\kappa_{v
\Psi \Psi}$ determines the relative strength. The spin $\dreih$
resonances are treated in the Rarita-Schwinger formalism as a pro\-duct
of a spin $1$ vector and a spin $\einh$ spinor, so in the coupling
term the $z_{\varphi \Psi}$ parameters determine the off shell
behavior of the resonance, which means the parameters affect the
appearance of the spin $\dreih$ resonances in spin $\einh$ channels.

The isospin factors for each vertex are listed in appendix
\ref{isofak}. These factors are chosen in a way to consistently
decompose the amplitudes into the different possible isospin channels.

Note that we do not introduce explicit background terms in our study
but instead generate the background consistently from the $u$- and
$t$-channel diagrams.

\subsection{Form Factors}
\label{ff}

As we are not dealing with-point like particles, we have to
introduce hadronic form factors in order to take care of the inner
structure of the particles which is important at the vertices.
The shape of the form factors is taken from \cite{fm} because our
goal is one consistent model for all meson- and photon-induced
reactions on the nucleon.

These shapes are given by the following functions:
\bea
F(q^2, m^2) =
\frac{\Lambda^4}{\Lambda^4 + (q^2 - m^2)^2}
\eea
in $s$- and $u$- channel, where $q^2 = s$ or $q^2 = u$ resp., and
\bea
F_t(t, m^2) = \frac{\Lambda^4 + ((t_{thresh} - m^2)/2)^2}
{\Lambda^4 + (t - (t_{thresh} + m^2)/2)^2}
\eea 
in $t$-channel diagrams. Here $t_{thresh}$ is the value of $t$ at
threshold. $m$ is the mass of the propagating particle and $\Lambda$
the form factor parameter. Note that purely mesonic vertices are not
multiplied by a form factor, i.e. the $t$-channel diagrams contain
only one form factor.

There are four different form factor parameters $\Lambda$, one for all
$t$-channels, one for all spin $\einh$ resonances, one for the Spin
$\dreih$ resonances and the $\Lambda$ parameter for diagrams with a
propagating asymptotic baryon (Born diagrams) which is taken from
\cite{fm}.

%%%%%%%%%%%%%%%%%%%%%%%%%%%%%%%%%%%%%%%%%%%%%%%%%%%%%%%%%%%%%%%%%%%%%%%%%
%
% Experimental Data
%
%%%%%%%%%%%%%%%%%%%%%%%%%%%%%%%%%%%%%%%%%%%%%%%%%%%%%%%%%%%%%%%%%%%%%%%%%

\section{Experimental Data}
\label{ex}

In this model we need to adjust parameters like the resonance masses and
widths to experimental data. Thus the accuracy of the extracted
parameters obviously depends on the quality of the available
experimental data. All experiments concerned with strangeness $-1$
resonances were carried out more than twenty years ago. The error bars
are quite large, and different experiments are often contradictory.

We are mainly using a partial wave analysis (PWA) \cite{gopal77} which
included a large amount of data and was performed for the three
channels we calculate ($\overline K N$, $\pi \Sigma$, $\pi
\Lambda$). This is an energy dependent PWA, which means that there
were already some assumptions made about resonance properties;
furthermore there is no error given in \cite{gopal77}. We thus have to
add some realistic error in order to use \cite{gopal77} together with
other data. The errors we use are
\be
\Delta T = \mathrm{Max}(0.12 \;T ;\; 0.012).
\ee
Almost all other PWAs (e.g. \cite{lw72},
\cite{alst78a}) are single channel analyses for the $K N$ channel. As
there are large deviations comparing the PWA of \cite{gopal77} to the
one of \cite{alst78a} (especially in the $S_{01}$ channel) we do not
want to take \cite{gopal77} as the only source of experimental
information. In addition, therefore, we use total cross sections for
$\pi \Sigma$ and $\pi \Lambda$ final states which are taken from
\cite{lb88} (because all experiments were done before \cite{lb88} was
published, this data collection can be considered complete for total
cross sections). For the $\overline K N$ channel there are also
differential cross sections available, so we are using the
publications given in 
table \ref{data}. All cross sections are measured with the initial
state $K^- p$. There are also data for $\knb n$ reactions, but for
those the error bars are even larger. 

With this set of experimental data we adjust the parameters of our
model (table \ref{par}). Since spin $\frac{5}{2}$ resonances are not
included in our calculation we do not go beyond $E_{cm} = 1.72$ GeV,
because at higher energies we expect a strong influence of the
resonances $\Lambda(1820)$, $\Lambda(1830)$, $\Sigma(1775)$, and
$\Sigma(1915)$ on cross sections. A reliable extraction of spin
$\dreih$ and $\einh$ resonance properties would not be possible if
contributions of these states were present in the data.

%%%%%%%%%%%%%%%%%%%%%%%%%%%%%%%%%%%%%%%%%%%%%%%%%%%%%%%%%%%%%%%%%%%%%%%%%
%
% Results
%
%%%%%%%%%%%%%%%%%%%%%%%%%%%%%%%%%%%%%%%%%%%%%%%%%%%%%%%%%%%%%%%%%%%%%%%%%

\section{Results}
\label{res}

In this section we show the partial wave amplitudes and differential
cross sections, calculated with the parameter set that yields the
least $\chi^2$ in the fit.

\subsection{Partial Waves}
Following \cite{pdb98} the nomenclature differs from the
non-strange case: $L_{I,2J}$ labels the amplitude with angular
momentum $L$, isospin $I$ and total spin $J$.

\begin{figure}
\begin{center}
\includegraphics[width=12cm]{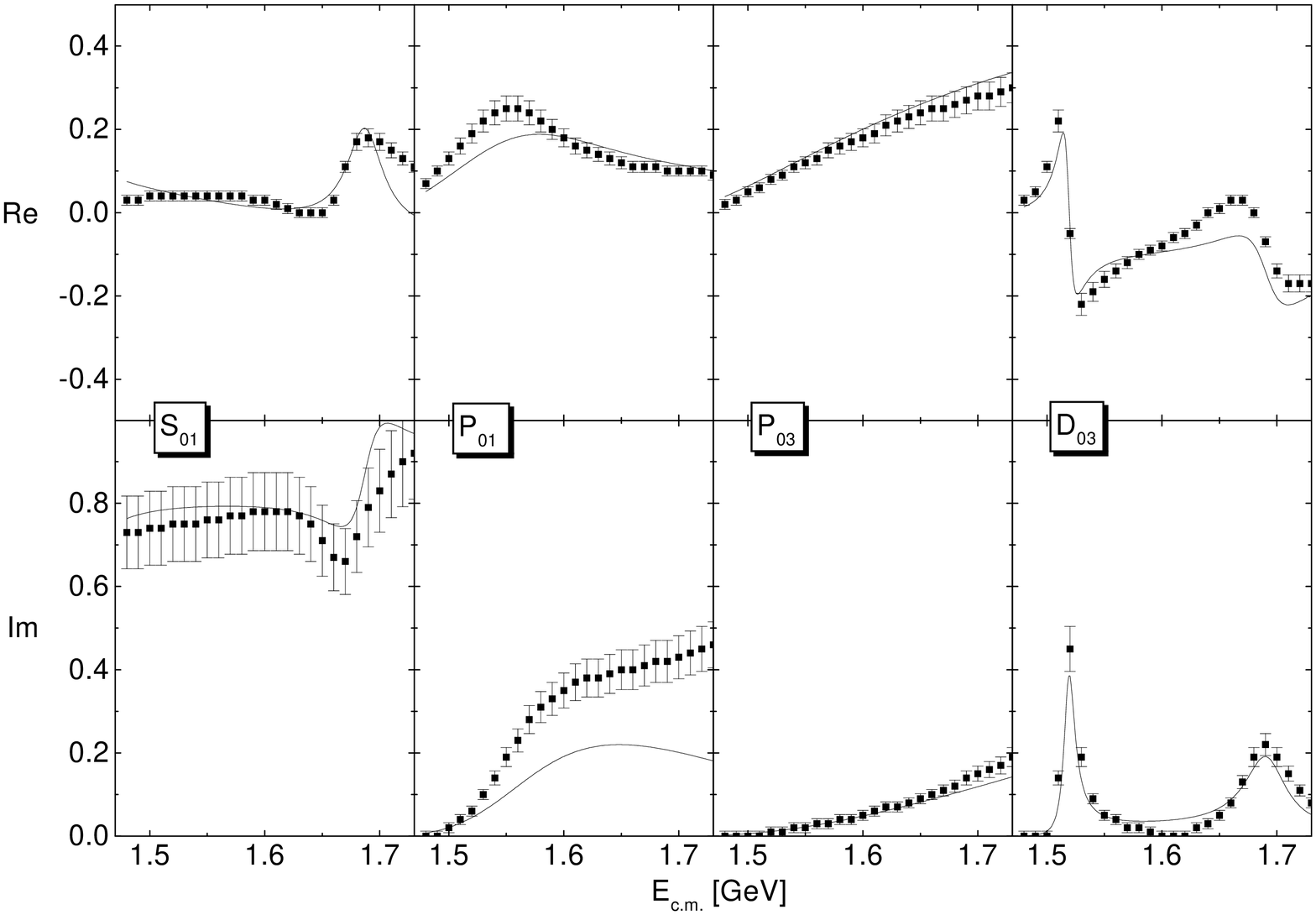} 
\end{center}
 \caption[PWA by \protect\cite{gopal77} and our calculation
 for $I=0$.]{PWA \protect\cite{gopal77} and our calculation
 for $K N \rightarrow K N $, $I=0$.}
\label{pwa_kn_i0}
\end{figure}

As already mentioned in section \ref{ex}, there are a number of
uncertainties in the analysis \cite{gopal77}. For example, in the
$S_{01}$ amplitude (fig. \ref{pwa_kn_i0}) the authors of
\cite{gopal77} had problems fixing the $KN$ width of the
$\Lambda(1670)$, which explains the deviations from our calculation.

The situation for the $P_{01}$ amplitude is not quite clear. There
might be one or two resonances in this partial wave \cite{pdb98}. In
our calculation we take only one resonance into account, since a
second resonance does not decrease the $\chi^2$ significantly. It is
difficult to disentangle background and resonant effects in the
available data; more accurate data would help clarifying this
situation.

No resonant contribution with quantum numbers $P_{03}$ is found, which
is consistent with \cite{pdb98}. The background contributions alone
already describe this partial wave amplitude very well.

The two established resonances $\Lambda(1520)$ and $\Lambda(1690)$ are
clearly seen in the $D_{03}$ amplitude where the fit works quite
well.

In the considered energy range there is no resonant contribution to
the $S_{11}$ amplitude. Including the $\Sigma(1750)$ might improve the
fit, but the parameters are not well known and it is not possible to
determine them in a fit constrained to energies below $E_{cm} = 1.72$ GeV.

The $P_{11}$ amplitude is in nice agreement with \cite{gopal77}. Only
for the $\pi \Lambda$ final state there are some discrepancies, but
this is clear since the $\Sigma(1660)$ was not included in the $\pi
\Lambda$ channel of \cite{gopal77}.

Describing the $P_{13}$ amplitude is difficult, because there is no
resonance in this partial wave. In addition the magnitude is quite
small, so it does not contribute much to the cross sections.

Our description of the $D_{13}$ is in good agreement with
\cite{gopal77}, only for the $\pi \Lambda$ final state the PWA by
\cite{gopal77} has a slightly different mass for the $\Sigma(1670)$
resonance.

It would be very helpful to have an energy independent PWA of
experimental data in these channels, since all energy dependent
analyses already contain some assumptions about resonance
properties.

\subsection{Cross Sections}

\begin{figure}[ht]
\begin{center}
\includegraphics[width=11.5cm]{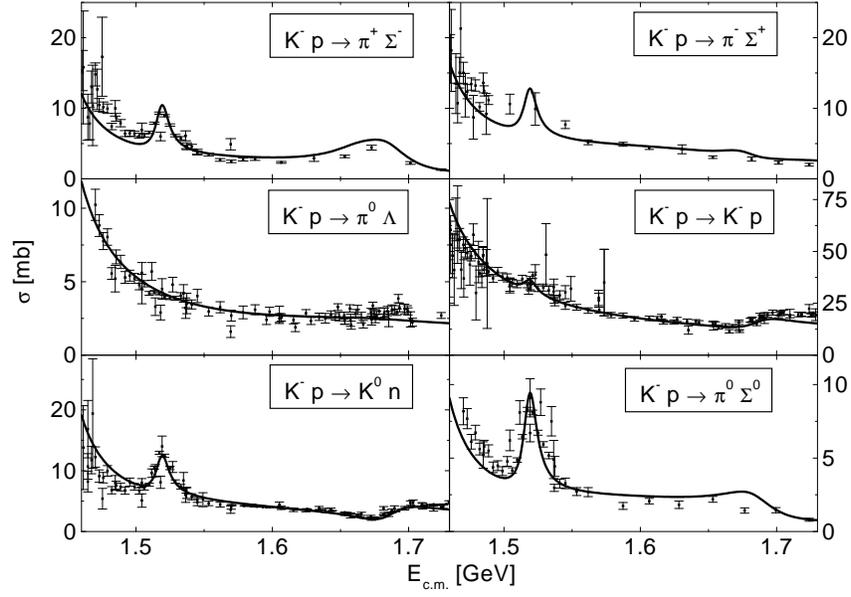} 
\end{center}
 \caption{Comparison of total cross sections for various reactions
   with data from \protect\cite{lb88} used in the fit.}
\label{tot_com}
\end{figure}

\begin{figure}[ht]
\begin{center}
\includegraphics[width=11.5cm]{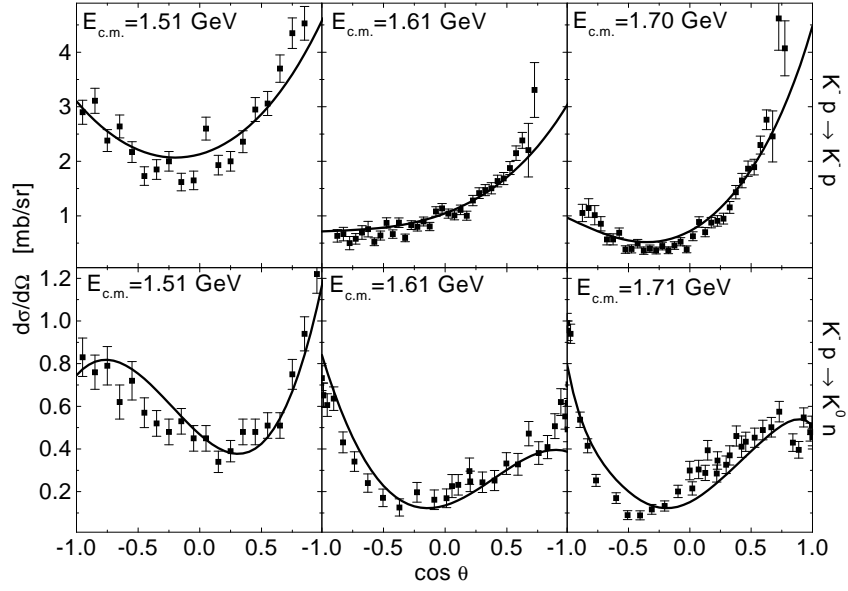} 
\end{center}
 \caption{The upper three plots show differential cross sections for
   the reaction $K^- p \rightarrow K^- p$, the lower ones for the
   reaction $K^- p \rightarrow \knb n$. The data are taken from the
   references given in table \ref{data}.}
\label{dif_com}
\end{figure}

We show a comparison of calculated total (fig. \ref{tot_com}) and
differential cross sections (fig. \ref{dif_com}) with the experimental
data used in the fit. The agreement is good within the errorbars but
it can also be seen in fig. \ref{tot_com} that the experimental data
are somewhat contradictory (especially for $K^- p \rightarrow \pi^0
\Lambda$).

\begin{figure}[ht]
\begin{center}
\includegraphics[width=11.5cm]{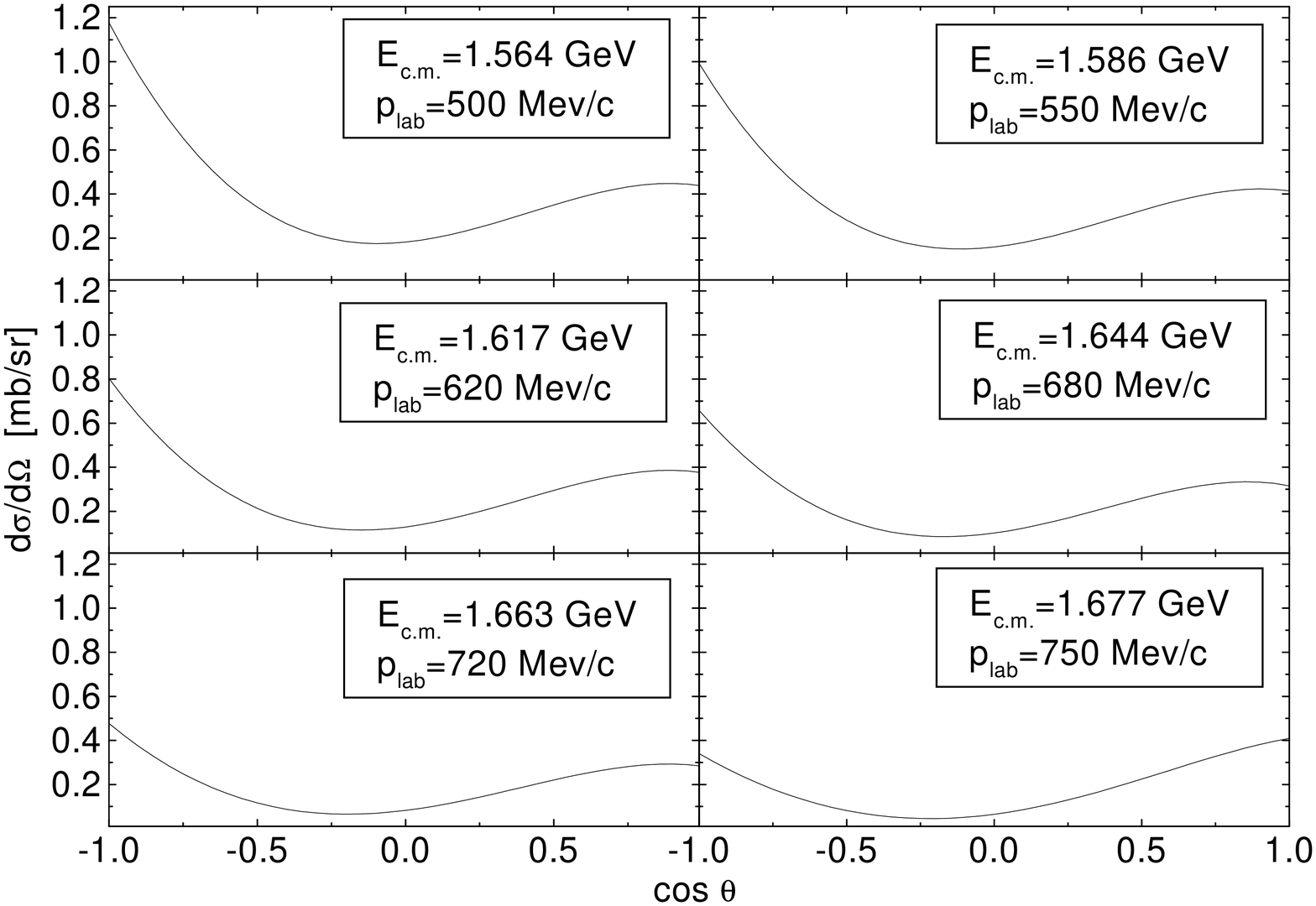} 
\end{center}
 \caption{Differential cross sections for the reaction $K^- p
   \rightarrow \knb n$.}
\label{dif_kn}
\end{figure}

\begin{figure}[ht]
\begin{center}
\includegraphics[width=11.5cm]{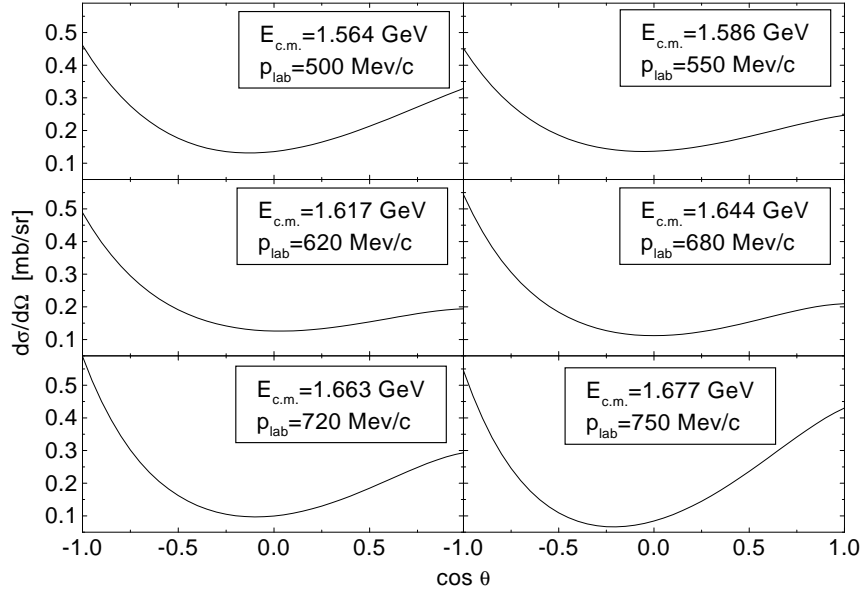} 
\end{center}
 \caption{Differential cross sections for the reaction $K^- p
   \rightarrow \pi^0 \Sigma^0$.}
\label{dif_ps}
\end{figure}

\begin{figure}[ht]
\begin{center}
\includegraphics[width=11.5cm]{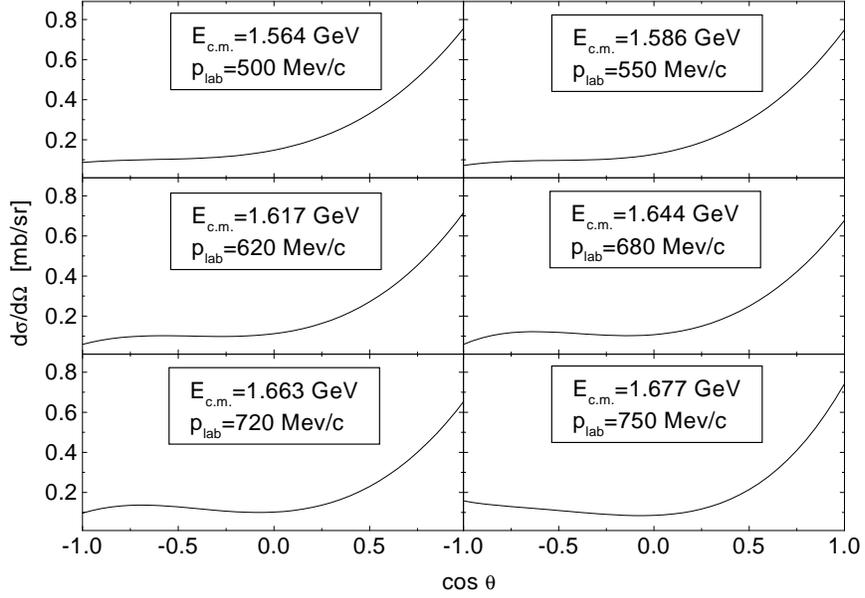} 
\end{center}
 \caption{Differential cross sections for the reaction $K^- p
   \rightarrow \pi^0 \Lambda $.}
\label{dif_pnl}
\end{figure}

In figs. \ref{dif_kn} through \ref{dif_pnl} we show as predictions the
calculated differential cross sections for the reactions $K^- p
\rightarrow \knb n$, $K^- p \rightarrow \pi^0 \Sigma^0$ and $K^- p
\rightarrow \pi^0 \Lambda $. The energies are chosen in view of the
upcoming data from the Crystal Ball collaboration \cite{bn00}.

\subsection{Coupling Constants}
\label{cou}

During the fit we adjusted the coupling constants to experimental
data. The couplings to final states are given in table \ref{ct} in
comparison with refs. \cite{KDOL00} and \cite{bh99}. The numbers from
\cite{KDOL00} are calculated within a sum rule approach with $SU(3)$
breaking and should be considered as a qualitative guide. The ones
from \cite{bh99} are calculated within a parametrisation of QCD that
goes back to \cite{mor89}, within this approach these numbers should
have an accuracy of about 10\%. Concrete values of coupling constants
are hard to find in the literature and show large deviations. All the
coupling constants that we obtained in our parametrisation from
fitting to experimental data have to be seen within the framework of
this $K$ matrix calculation. Since the experimental data have error
bars larger than 10\% the extracted parameters cannot be more
accurate. In addition, with only total cross sections for the $\pi
\Sigma$ and $\pi \Lambda$ final states it is not possible to
disentangle resonant and different background contributions (see also
the comments on spin $\einh$ resonances in the following section).

Comparing the $F/D$ ratios (as defined in \cite{pdb98}) that fit our
couplings show that these couplings are not $SU(3)$ symmetric. The
actual values are of minor importance.

\subsection{Resonances}

In tables \ref{massvergl} and \ref{parvergl} we show the extracted
properties of all the included resonances. The masses for
$\Lambda(1405)$ and $\Sigma(1385)$ were taken from \cite{pdb98},
because they are rather accurately known and cannot be determined in
our fit due to their values below the $\overline K N$ threshold. 

The values that are labeled '$K$ matrix' are the ones that are used in
the calculation. The resonance widths are used to calculate the
coupling constants which enter the $K$ matrix.

With the help of the speed technique we extracted the resonance
parameters from our partial wave amplitudes. The concept is as follows
\cite{fm}, \cite{hoe93}: 

The ansatz for a resonance is given by a Breit-Wigner shape
(cf. \cite{pdb98}) 
\be 
T(W) = T_{back} (W) + 
\frac{\rmx \; \Gamma / 2 \; \rme^{\rmi \Phi}}{m_R - W - \rmi \Gamma/2} \;
,
\label{efftmat}
\ee
where $W = \sqrt{s}$, $m_R$ is the mass of the resonance, $\Gamma$ its
width and $\rmx = \sqrt{\Gamma_i\Gamma_f} / \Gamma$ for partial widths
$\Gamma_i$ and $\Gamma_f$ of the resonance's decay into initial and
final state. Under the assumption that the background ($T_{back}$)
can be considered constant compared to the resonant contributions in
the vicinity of the resonance, one gets
\be
\frac{dT}{dW} \approx \frac{\rmx \; \Gamma / 2 \rme^{\rmi \Phi}}
{(m_R - W - \rmi \Gamma/2)^2}\; .
\ee
The speed is defined by
\be
Sp(W) = \abs {\frac{dT}{dW}} \approx \frac{\rmx \; \Gamma / 2}{(m_R -
  W)^2 + \Gamma^2/4} \; . 
\label{speedan}
\ee
Fitting the parameters of the Breit-Wigner shape to the calculated
speed we can extract the desired properties of the resonances.

From tables \ref{massvergl} and \ref{parvergl} it can be seen that the
parameters extracted via the speed technique and the $K$ matrix
parameters are almost equal for spin $\dreih$ resonances. This is due
to the small width of all these resonances and the fact that there is
not much background in these partial wave amplitudes. All these
parameters are in very good agreement with the values given in
\cite{pdb98}.

\begin{figure}[ht]
\begin{center}
\includegraphics[width=12cm]{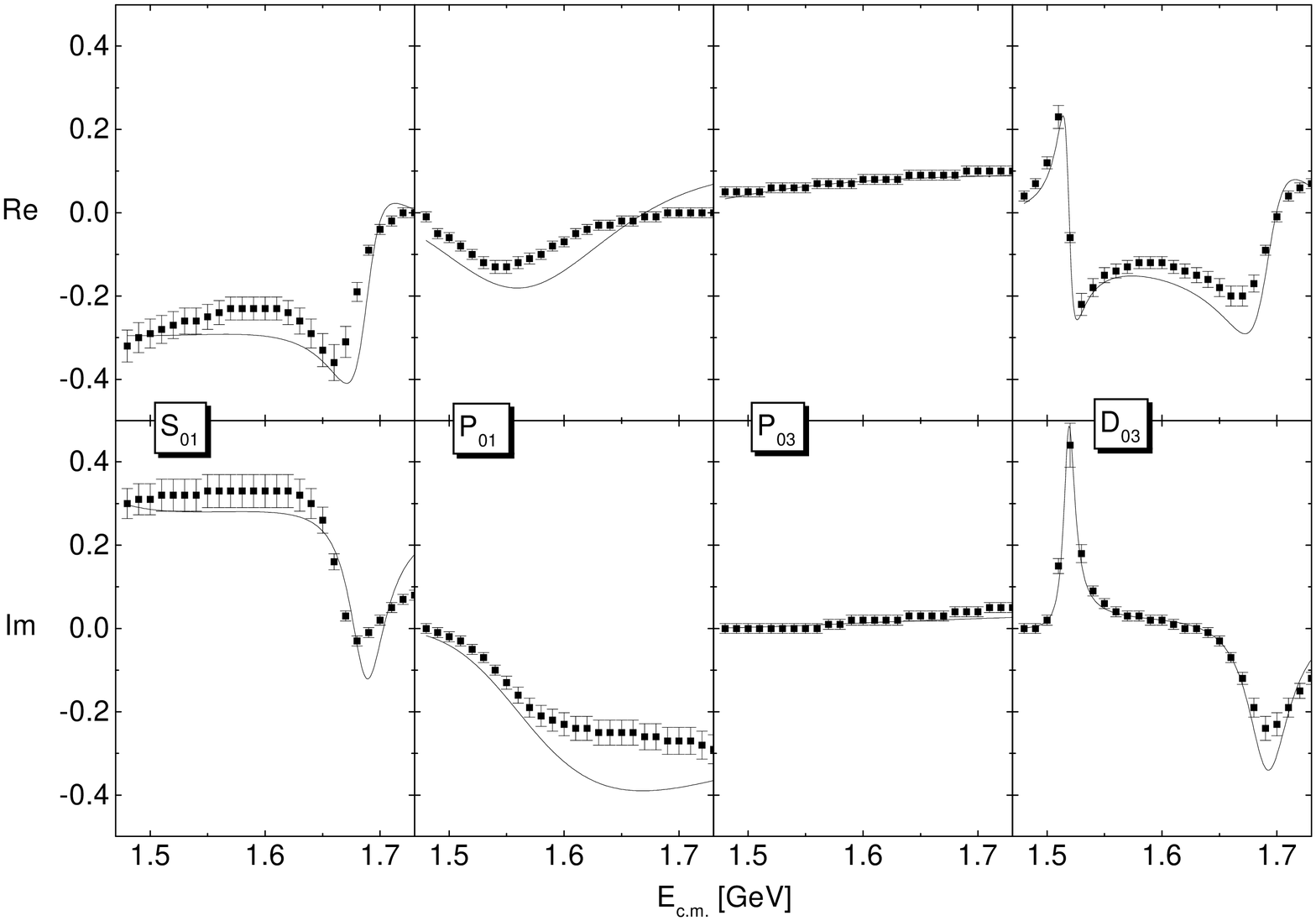} 
\end{center}
 \caption[PWA by \protect\cite{gopal77} and our calculation
 for $I=0$.]{PWA \protect\cite{gopal77} and our calculation
 for $K N \rightarrow \pi \Sigma $, $I=0$.}
\label{pwa_ps_i0}
\end{figure}

The situation for the spin $\einh$ resonances is completely
different. There are large background contributions in these
amplitudes. This is most dramatically the case for the $P_{01}$
$\Lambda (1600)$ resonance for which the $K$ matrix analysis gives a
pole mass of 1710 MeV whereas the speed analysis yields 1545 MeV. In
this case, the speed technique is not applicable, since the background
varies more than the resonant part of the amplitude. A
similar problem occurs for the $S_{01}$ $\Lambda (1670)$. Here the
speed technique seems to show a quite narrow resonance with a width of
52 MeV, but the actual decay width, calculated from the coupling in
the $K$ matrix, is almost 960 MeV. The careful examination of this 
puzzle shows, that off-shell contributions of the spin $\dreih$
resonances $\Lambda(1520)$ and $\Lambda(1690)$ together with resonant
contributions from the very broad $\Lambda(1670)$ add up to the
$S_{01}$ amplitude that is given in figs \ref{pwa_kn_i0},
\ref{pwa_ps_i0} yielding a quite sharp peak in the speed plot. This
example just shows that the definition of resonance properties by a
Breit-Wigner shape in the speed technique may be quite different from
the particle properties obtained from the $K$ matrix parameters.

This interplay of different contributions to a resonance-like
structure in the amplitudes makes it hard to get a unique set of
parameters, which describes the experimental data. With the
available data it is not possible to disentangle the necessary
contributions from Born-, $u$-channel, and resonant diagrams. In
addition, the off shell parameters of the spin 
$\dreih$ resonances play an important role. Better experimental data
would surely lead to a more unique set of parameters in this
field. The overall description is not bad, comparing the speed results
to the experimental ones.

%%%%%%%%%%%%%%%%%%%%%%%%%%%%%%%%%%%%%%%%%%%%%%%%%%%%%%%%%%%%%%%%%%%%%%%%%
%
% Confusion and Outlook
%
%%%%%%%%%%%%%%%%%%%%%%%%%%%%%%%%%%%%%%%%%%%%%%%%%%%%%%%%%%%%%%%%%%%%%%%%%

\section{Conclusions \& Outlook}
\label{cno}

We have presented the first results of our calculations for $\overline
K$ induced reactions on the nucleon. The model of \cite{fm} has been
extended to the strangeness $-1$ sector. All available data are
described quite well within the given error bars. The extracted
resonance masses and widths agree with the ones given in
\cite{pdb98}. Since the accuracy of the known parameters is not very
high there is a lot of room for improvement. Some of the resonances
are not established at all and will not be established before better
experimental data are published.

Putting the calculations for $\gamma$, $\pi$ and $\overline K$ induced
reactions on the nucleon together we have developed a consistent way
of treating all hadronic reactions on the nucleon below $1.72$
GeV. Currently we are working on implementing the vector mesons $\rho$
and $\omega$ as asymptotic particles. As soon as there are better data
available in the strangeness sector we will also start implementing
more asymptotic states like e.g. $\eta \Lambda$ and $\pi \pi \Sigma$.

Including spin $\frac{5}{2}$ resonances would certainly also improve
the calculations and extend the accessible energy range, but here the
problem of a consistent treatment is not solved up to now.

\section*{Acknowledgements}

We are grateful to B. Nefkens for extensive and helpful discussion on
the Crystal Ball data. Special thanks go to T. Feuster for the
original $\pi N$ code which was modified for our calculations.

\clearpage
%%%%%%%%%%%%%%%%%%%%%%%%%%%%%%%%%%%%%%%%%%%%%%%%%%%%%%%%%%%%%%%%%%%%%%%%%
%
% Appendix Isospin Factors
%
%%%%%%%%%%%%%%%%%%%%%%%%%%%%%%%%%%%%%%%%%%%%%%%%%%%%%%%%%%%%%%%%%%%%%%%%%
\appendix

\section{Isospin Factors}
\label{isofak}

The listed factors are contained in the Lagrange density and thus
enter the amplitude at each vertex. Outgoing particles are noted with
a $\overline{bar}$, so the outgoing $\knb$ reads $\overline{\knb}$.

\bigskip

The $\Lambda K N$ vertex:
\bea
g_{\Lambda K N} (- \mb \Lambda p K^- + \mb \Lambda n \knb - \mb
K^- \mb p \Lambda + \mb \knb \mb n \Lambda)
\eea

\bigskip

The $\Sigma K N$ vertex:
\bea
g_{\Sigma K N} (\sqrt 2 \; \mb \Sigma^+\knb p + \mb \Sigma^0 K^- p +
\mb \Sigma^0 \knb n + \sqrt 2 \; \mb \Sigma^- K^- n \nonumber
\\ 
+ \sqrt 2 \; \mb \knb \mb p \mb \Sigma^+ + \mb K^- \mb p \Sigma^0 +
\mb \knb \mb n \Sigma^0 + \sqrt 2 \; \mb K^- \mb n \Sigma^-)
\eea

\bigskip

The $\Lambda \pi \Sigma$ vertex:
\bea
g_{\Lambda \pi \Sigma} (\mb \Lambda \pi^+ \Sigma^- - \mb \Lambda \pi^0
\Sigma^0 + \mb \Sigma^- \mb \pi^+ \Lambda - \mb \Sigma^0 \mb \pi^0
\Lambda + \mb \Sigma^+ \mb \pi^- \Lambda)
\eea 

\bigskip

The $\Sigma \pi \Sigma$ vertex:
\bea
g_{\Sigma \pi \Sigma} (
\mb \Sigma^- \pi^0 \Sigma^- - \mb \Sigma^- \pi^- \Sigma^0 + 
\mb \Sigma^0 \pi^+ \Sigma^- - \mb \Sigma^0 \pi^- \Sigma^+ + 
\mb \Sigma^+ \pi^+ \Sigma^0 - \mb \Sigma^+ \pi^0 \Sigma^+ 
\nonumber \\
+ \mb \Sigma^- \mb \pi^0 \Sigma^- - \mb \Sigma^0 \mb \pi^- \Sigma^- +
\mb \Sigma^- \mb \pi^+ \Sigma^0 - \mb \Sigma^+ \mb \pi^- \Sigma^0 +
\mb \Sigma^0 \mb \pi^+ \Sigma^+ - \mb \Sigma^+ \mb \pi^0 \Sigma^+)
\eea

\bigskip

The $\Delta \pi N$ vertex:
\bea
g_{\Delta \pi N} 
( \sqrt \eind \; \mb n \Delta^+ \pi^- 
- \sqrt \zweid \; \mb n \Delta^0 \pi^0  + \mb n \Delta^- \pi^+
\nonumber \\
+ \mb p \Delta^{++} \pi^- - \sqrt \zweid \; \mb p \Delta^+ \pi^0
+ \sqrt \eind \; \mb p \Delta^0 \pi^+ \nonumber
\\
+ \sqrt \eind \;\mb \Delta^+ \mb \pi^- n
- \sqrt \zweid \;\mb \Delta^0 \mb \pi^0 n + \mb  \Delta^- \mb \pi^+ n
\nonumber \\
+ \mb \Delta^{++} \mb \pi^- p - \sqrt \zweid \; \mb \Delta^+ \mb \pi^0 p
+ \sqrt \eind \; \mb \Delta^0 \mb \pi^+ p)
\eea

\bigskip

The $N \pi N$ vertex:
\bea
g_{N \pi N} 
(\mb n \pi^0 n - \sqrt 2 \; \mb n \pi^- p + \sqrt 2 \; \mb p \pi^+ n -
\mb p \pi^0 p \nonumber 
\\
\mb n \; \mb \pi^0 n - \sqrt 2 \; \mb p \; \mb \pi^- n + \sqrt 2 \;
\mb n \; \mb \pi^+ p - \mb p \; \mb \pi^0 p)
\eea

\bigskip

The $\Sigma K \Delta$ vertex:
\bea 
g_{\Sigma K \Delta}
( \mb \Sigma^+ K^- \Delta^{++} - \sqrt \eind \; \mb \Sigma^+ \knb
\Delta^+ + \sqrt \zweid \; \mb \Sigma^0 K^- \Delta^+ \nonumber
\\
- \sqrt \zweid \; \mb \Sigma^0 \knb \Delta^0 + \sqrt \eind \; \mb
\Sigma^- \knb \Delta^- - \mb \Sigma^- \knb \Delta^- \nonumber 
\\
\mb K^- \mb \Delta^{++} \Sigma^+ - \sqrt \eind \; \mb \knb \mb
\Delta^+ \Sigma^+ + \sqrt \zweid \; \mb K^- \mb \Delta^+ \Sigma^0
\nonumber \\
- \sqrt \zweid \; \mb \knb \mb \Delta^0 \Sigma^0 + \sqrt \eind \; \mb
\knb \mb \Delta^- \Sigma^- - \mb \knb \mb \Delta^- \Sigma^-)
\eea

\clearpage
%%%%%%%%%%%%%%%%%%%%%%%%%%%%%%%%%%%%%%%%%%%%%%%%%%%%%%%%%%%%%%%%%%%%%%%%%
%
% Bibliography
%
%%%%%%%%%%%%%%%%%%%%%%%%%%%%%%%%%%%%%%%%%%%%%%%%%%%%%%%%%%%%%%%%%%%%%%%%%

\bibliographystyle{unsrt}

\newpage

%%%%%%%%%%%%%%%%%%%%%%%%%%%%%%%%%%%%%%%%%%%%%%%%%%%%%%%%%%%%%%%%%%%%%%%%%
%
% tables
%
%%%%%%%%%%%%%%%%%%%%%%%%%%%%%%%%%%%%%%%%%%%%%%%%%%%%%%%%%%%%%%%%%%%%%%%%%

\begin{table}[ht]
\begin{center}
  \begin{tabular}{lll}
    \hline
    Reactions                   & Energy Range        & References   \\
    \hline
    $ K^- p \rightarrow K^- p$ & $1.48$ to 1.55 GeV & \cite{mast76}\\
                               & $1.61$ to 1.72 GeV & \cite{adams75}\\
    \hline
    $ K^- p \rightarrow\knb p$ & $1.49$ to 1.54 GeV & \cite{mast76}\\
                               & $1.57$ to 1.72 GeV & \cite{alst78}\\
    \hline
  \end{tabular}
\end{center}
\caption{Differential cross sections used in the fit}
\label{data}
\end{table}

\begin{table}[ht]
\begin{center}
\begin{tabular}{llr}
\hline
\multicolumn{3}{c}{38 Resonance Parameters}\\
\hline
$\Lambda$-Resonances & Masses                               & 4\\
(1405), (1520), (1600), & $K N$-Width                 & 5 \\
(1670), (1690)          & $\pi \Sigma$-Width                   & 5 \\
                       & $z_{K N}$ (for Spin $\dreih$)        & 2 \\
                       & $z_{\pi \Sigma}$ (for Spin $\dreih$) & 2 \\
\hline
$\Sigma$-Resonances & Masses                                & 2 \\
(1385), (1660), (1670) & $K N$-Width                           & 3 \\
                    & $\pi \Sigma$-Width                    & 3 \\
                    & $\pi \Lambda$-Width                   & 3 \\
                    & $z_{K N}$ (for Spin $\dreih$)         & 3 \\
                    & $z_{\pi \Sigma}$ (for Spin $\dreih$)  & 3 \\
                    & $z_{\pi \Lambda}$ (for Spin $\dreih$) & 3 \\
\hline
\multicolumn{3}{c}{15 Background Couplings} \\
\hline
Couplings  & & \\
-- of Born Terms     & $g_{K N \Sigma}$, $g_{K N \Lambda}$,  $g_{\pi
  \Sigma \Sigma}$, $g_{\pi \Sigma \Lambda}$ & 4\\
-- of t-channels      & $g_{K^* N \Lambda}$, $g_{K^* N \Sigma}$,
$g_{a_0 N N}$, $g_{\Phi N N}$, $g_{\rho \Lambda \Sigma}$, $g_{\rho
  \Sigma \Sigma}$ & 6 \\ 
                     &  $\kappa_{K^* N \Lambda}$, $\kappa_{K^* N
                       \Sigma}$, $\kappa_{\Phi N
                       N}$, $\kappa_{\rho \Lambda \Sigma}$,
                     $\kappa_{\rho \Sigma \Sigma}$ & 5 \\
\hline
\multicolumn{3}{c}{3 Form Factor Parameters} \\
\hline
One $\Lambda$ Parameter for all & & \\
-- Spin-$\einh$ Resonances  & $\Lambda_\einh$ & 1 \\
-- Spin-$\dreih$ Resonances & $\Lambda_\dreih$ & 1 \\
-- $t$-channel              & $\Lambda_{v \Psi \Psi}$ & 1 \\
\hline
\end{tabular}
\end{center}
\caption{Model parameters}
\label{par}
\end{table}

\begin{table}[ht]
\begin{center}
\begin{tabular}{lccccc}
\hline
& $|g_{\pi N N}|$ & $|g_{KN\Lambda}|$ & $|g_{KN\Sigma}|$ &
$|g_{\pi\Sigma\Lambda}|$ & $|g_{\pi\Sigma\Sigma}|$ \\
\hline
this calculation & 1 & 0.71 & 0.16 & 0.40 & 0.04 \\
\hline
\cite{KDOL00}           & 1 & --- & --- & --- & 0.27 \\
\hline
\cite{bh99} & 1 & --- & --- & 0.82 & 0.54 \\
\hline
\hline
$F/D$ & --- & 0.14 & 1.38 & 1.94 & 0.02 \\
\hline
\end{tabular}
\end{center}
\caption{Extracted coupling constants in comparison to other authors
  and $F/D$ ratio.}
\label{ct}
\end{table}

\begin{table}[ht!]
\begin{center}
\begin{tabular}{lllc}
\hline
\multicolumn{4}{c}{$\Lambda$-Resonances}\\
\hline
                  & \multicolumn{3}{c}{Mass [MeV]} \\
\hline
                  & $K$ Matrix & Speed         & \cite{pdb98} \\
\hline
$\Lambda(1405) \; \; S{01}$   & 1406.0 &  ---   & 1406.5 $\pm$ 4.0 \\
$\Lambda(1520) \; \; D{03}$   & 1519.3 & 1518.6 & 1519.5 $\pm$ 1.0 \\
$\Lambda(1600) \; \; P{01}$   & 1710.0 & 1545.0$^1$ & 1560 to 1700  \\
$\Lambda(1670) \; \; S{01}$   & 1727.4 & 1687.3$^1$ & 1660 to 1680 \\
$\Lambda(1690) \; \; D{03}$   & 1694.6 & 1691.3 & 1685 to 1695 \\
\hline
\hline
\multicolumn{4}{c}{$\Sigma$-Resonances}\\
\hline
                  & \multicolumn{3}{c}{Mass [MeV]} \\
\hline
                  & $K$ Matrix & Speed         & \cite{pdb98} \\
\hline
$\Sigma(1385) \; \; P{13}$   & 1383.0 &   ---  & 1382.8 $\pm$ 0.4 \\
$\Sigma(1660) \; \; P{11}$   & 1743.0 & 1661.6$^1$& 1630 to 1690 \\
$\Sigma(1670) \; \; D{13}$   & 1671.4 & 1665.8 & 1665 to 1685 \\
\hline
\end{tabular}
\end{center}
\caption{Comparison of the resonance masses from our $K$ matrix,
  extracted with the help of the speed-technique and the values given
  in \protect\cite{pdb98}. See text for the uncertainties
  concerning the numbers labeled $^1$.} 
\label{massvergl}
\end{table}

\begin{table}[ht!]
\begin{center}
\begin{tabular}{llcccc}
\hline
\multicolumn{6}{c}{$\Lambda$-Resonances}\\
\hline
         &  & $\Gamma$ [MeV] & $\mathrm x_{KN}$ [\%] & $\mathrm
           x_{\pi\Sigma}$ [\%] &  \\
\hline
$\Lambda(1405)$ & $K$ Matrix   & 183.3            &  ---           & 100 & \\
                & Speed        & ---              & ---            & --- & \\
                & \cite{pdb98} &  50 $\pm$ 2      & ---         & 100 & \\
\hline
$\Lambda(1520)$ & $K$ Matrix     & 11.9           & 39         & 61 & \\
                & Speed          & 11.7           & 37.9       & 60.5 & \\
                & \cite{pdb98}   & 15.6 $\pm$ 1.0 & 45 $\pm$ 1 & 42 $\pm$ 1 &\\
\hline
$\Lambda(1600)$ & $K$ Matrix     & 704.2            & 18             & 82       & \\
                & Speed          & 170$^1$  & 19.3$^1$           &   64$^1$   & \\
                & \cite{pdb98}   & 50 to 250       & 15 to 30      & 10 to 60  & \\
\hline
$\Lambda(1670)$ & $K$ Matrix     &  959.5            & 97             & 3          & \\
                & Speed          &  52$^1$  & 20$^1$ & 14$^1$  & \\
                & \cite{pdb98}   &  25 to 50       & 15 to 25      & 20 to 60  &  \\
\hline
$\Lambda(1690)$ & $K$ Matrix     & 45.4             & 15             & 85         & \\
                & Speed          & 45   & 15 & 38  & \\
                & \cite{pdb98}   & 50 to 60        & 20 to 30      & 20 to 40  &  \\
\hline
\hline
\multicolumn{6}{c}{$\Sigma$-Resonances}\\
\hline
         &  & $\Gamma$ [MeV] & $\mathrm x_{KN}$ [\%] & $\mathrm
           x_{\pi\Sigma}$ [\%] & $\mathrm x_{\pi\Lambda}$ [\%]  \\
\hline
$\Sigma(1385)$  & $K$ Matrix     & 15.1         & ---           & 1.7 & 13.4 \\
                & Speed          & ---          & ---            & --- & \\
                & \cite{pdb98}   & 36 $\pm$ 5   & ---            & 12 $\pm$ 2 & 88 $\pm$ 2 \\
\hline
$\Sigma(1660)$ & $K$ Matrix     & 481.2            & 4              & 77  & 19 \\
               & Speed          & 200$^1$  &  11$^1$ & ---$^2$  & 14$^1$ \\
               & \cite{pdb98}   & 40 to 200       & 10 to 30      & ---  & ---  \\
\hline
$\Sigma(1670)$  & $K$ Matrix     & 49.6             & 7             & 89   &  4 \\
                & Speed          & 54.9   & 9   & 26 & 5 \\
                & \cite{pdb98}   & 40 to 80        &  7 to 13      & 30 to 60 & 5 bis 15 \\
\hline
\end{tabular}
\end{center}
\caption{Comparison of the resonance parameters used in the $K$ matrix,
  extracted with the speed-technique, with the ones given in
  \protect\cite{pdb98}. See text for the uncertainties
  concerning the numbers labeled $^1$. Extraction failed for $^2$.} 
\label{parvergl}
\end{table}

%%%%%%%%%%%%%%%%%%%%%%%%%%%%%%%%%%%%%%%%%%%%%%%%%%%%%%%%%%%%%%%%%%%%%%%%%
%
% figures
%
%%%%%%%%%%%%%%%%%%%%%%%%%%%%%%%%%%%%%%%%%%%%%%%%%%%%%%%%%%%%%%%%%%%%%%%%%

\begin{figure}[ht]
\begin{center}
\includegraphics[width=12cm]{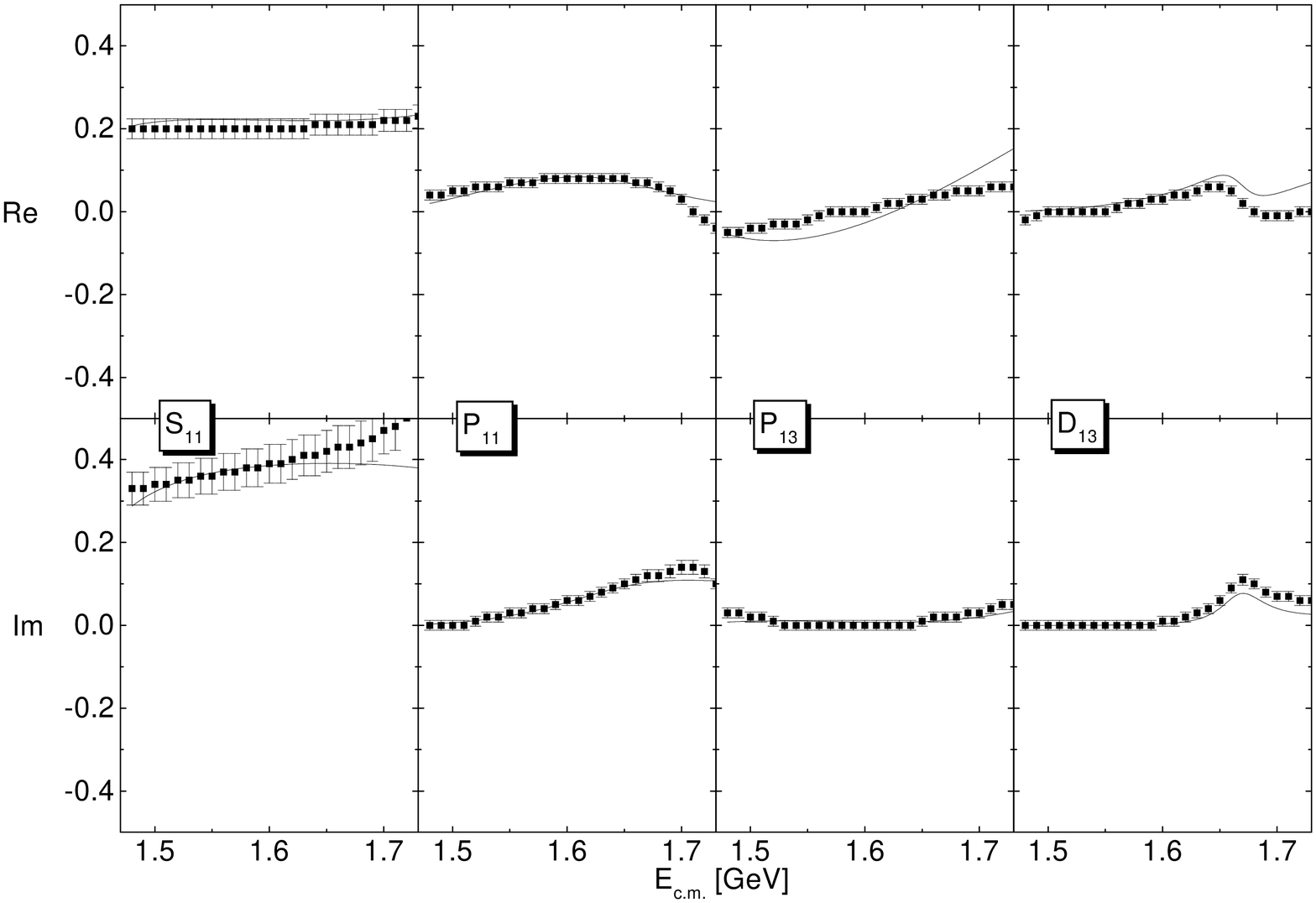} 
\end{center}
 \caption[PWA by \protect\cite{gopal77} and our calculation
 for $I=1$.]{PWA \protect\cite{gopal77} and our calculation
 for $K N \rightarrow K N $, $I=1$.}
\label{pwa_kn_i1}
\end{figure}

\begin{figure}[ht]
\begin{center}
\includegraphics[width=12cm]{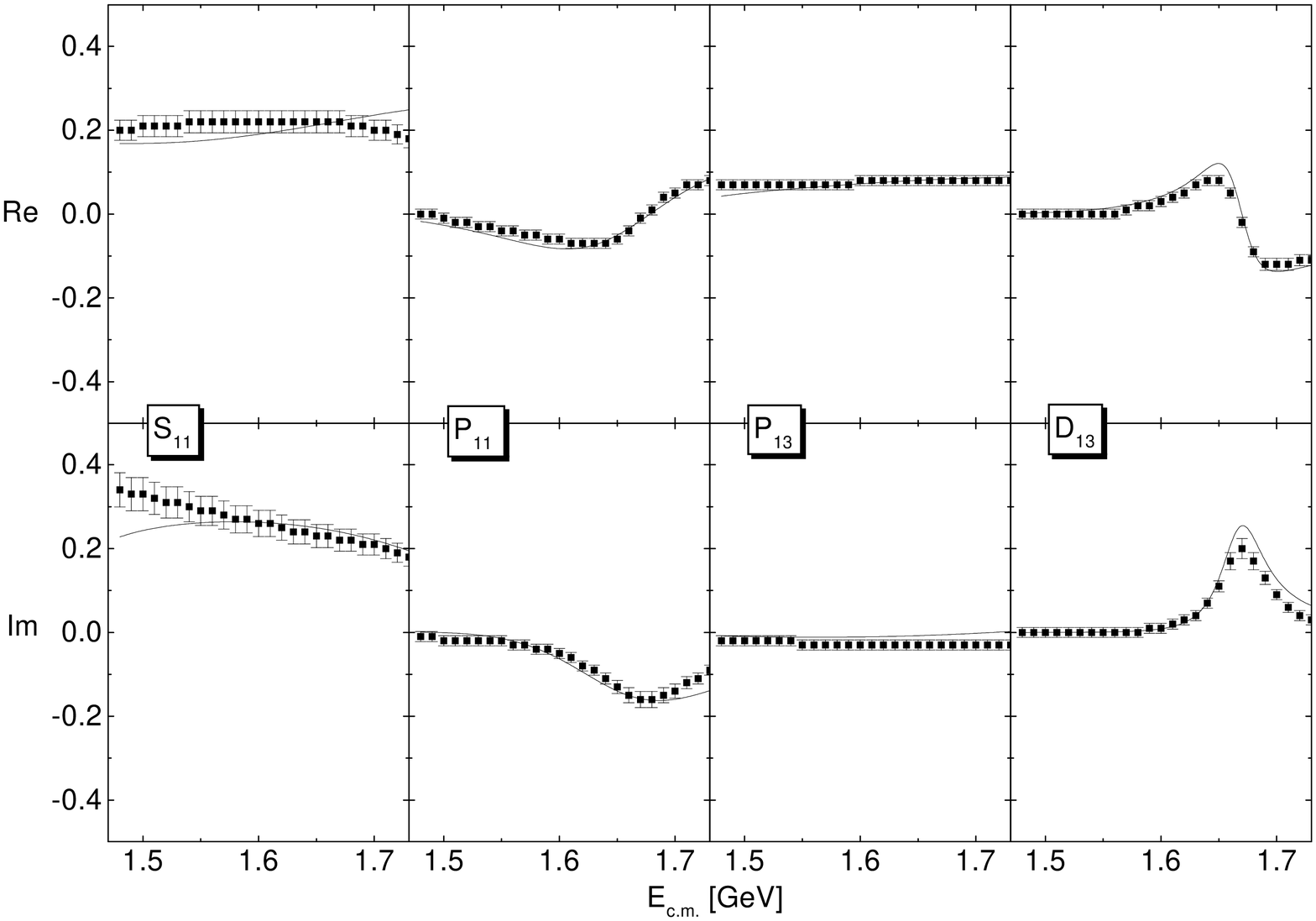} 
\end{center}
 \caption[PWA by \protect\cite{gopal77} and our calculation
 for $I=1$.]{PWA \protect\cite{gopal77} and our calculation
 for $K N \rightarrow \pi \Sigma $, $I=1$.}
\label{pwa_ps_i1}
\end{figure}

\begin{figure}[ht]
\begin{center}
\includegraphics[width=12cm]{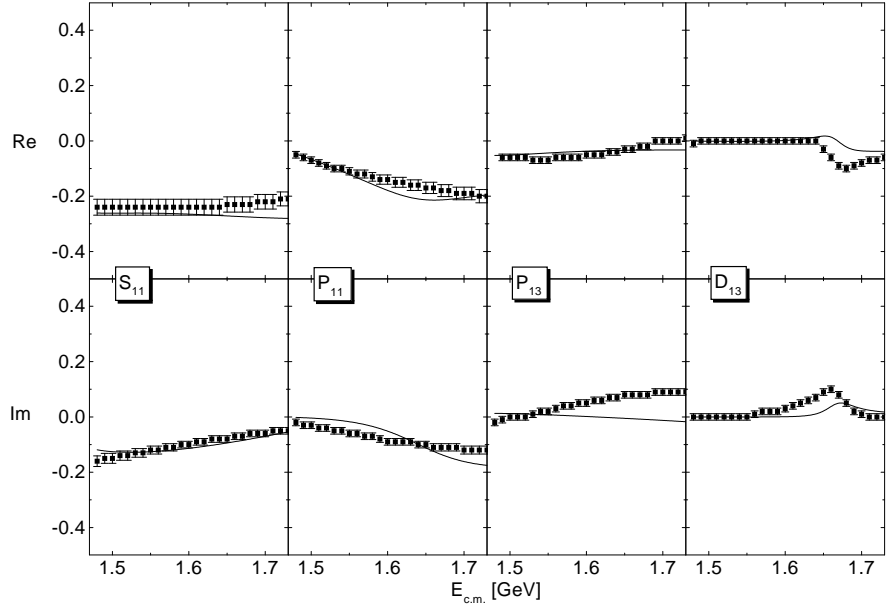} 
\end{center}
 \caption[PWA by \protect\cite{gopal77} and our calculation
 for $I=1$.]{PWA \protect\cite{gopal77} and our calculation
 for $K N \rightarrow \pi \Lambda $, $I=1$.}
\label{pwa_pl_i1}
\end{figure}

\clearpage

\end{document}